\documentclass[reprint, superscriptaddress, secnumarabic, amssymb, nobibnotes, aps, prl]{revtex4-1}

\usepackage{graphicx}
\usepackage{epstopdf}
\usepackage[T1]{fontenc}
\usepackage[latin9]{inputenc}
\usepackage{amsbsy}
\usepackage{gensymb}
\usepackage[T1]{fontenc}
\usepackage[latin9]{inputenc}
\usepackage{amsmath}
\usepackage{amssymb}
\usepackage{bbm}
\usepackage{braket}
\usepackage{xcolor}
\allowdisplaybreaks
\usepackage{graphicx}
\usepackage[colorlinks=true]{hyperref}  
\hypersetup{
    bookmarks=true,         % show bookmarks bar?
    unicode=false,          % non-Latin characters 
    pdftoolbar=true,        % show Acrobat
    pdfmenubar=true,        % show Acrobat 
    pdffitwindow=false,     % window fit to page when opened
    pdfstartview={FitH},    % fits the width of the page to the window
    pdftitle={HEA_2${nd}$},    % title
    pdfauthor={},     % author
    pdfsubject={},   % subject of the document
    pdfcreator={},   % creator of the document
    pdfproducer={}, % producer of the document
    pdfkeywords={} {} {}, % list of keywords
    pdfnewwindow=true,      % links in new window
    colorlinks=true,       % false: boxed links; true: colored links
    linkcolor=blue, %red,          % color of internal links (change box color with linkbordercolor)
    citecolor=blue,        % color of links to bibliography
    filecolor=magenta,      % color of file links
    urlcolor=blue           % color of external links
} 
\usepackage[normalem]{ulem}

\newcommand{\equref}[1]{Eq.~(\ref{#1})}

\newcommand{\figref}[1]{Fig.~\ref{#1}}

\renewcommand{\approx}{\simeq}

\begin{document}

\title{\textrm{Superconducting and normal state properties of  high entropy alloy Nb-Re-Hf-Zr-Ti investigated by muon spin relaxation and rotation}}
\author{Kapil Motla}
\affiliation{Department of Physics, Indian Institute of Science Education and Research Bhopal, Bhopal, 462066, India}
\author{Pavan Meena}
\affiliation{Department of Physics, Indian Institute of Science Education and Research Bhopal, Bhopal, 462066, India}
\author{Arushi}
\affiliation{ISIS Facility, STFC Rutherford Appleton Laboratory, Harwell Science and Innovation Campus, Oxfordshire, OX11 0QX, UK}
\author{D. Singh}
\affiliation{ISIS Facility, STFC Rutherford Appleton Laboratory, Harwell Science and Innovation Campus, Oxfordshire, OX11 0QX, UK}
\author{P. K. Biswas}
\affiliation{ISIS Facility, STFC Rutherford Appleton Laboratory, Harwell Science and Innovation Campus, Oxfordshire, OX11 0QX, UK}
\author{A. D. Hillier}
\affiliation{ISIS Facility, STFC Rutherford Appleton Laboratory, Harwell Science and Innovation Campus, Oxfordshire, OX11 0QX, UK}
\author{R. P. Singh}
\email[]{rpsingh@iiserb.ac.in}
\affiliation{Department of Physics, Indian Institute of Science Education and Research Bhopal, Bhopal, 462066, India}

\date{30-12-2020}
\begin{abstract}
\begin{flushleft}
\end{flushleft}
Superconducting high entropy alloy (HEA) are emerging as a new class of superconducting materials. It provides a unique opportunity to understand the complex interplay of disorder and superconductivity. We report the synthesis and detail bulk and microscopic characterization of Nb$_{60}$Re$_{10}$Zr$_{10}$Hf$_{10}$Ti$_{10}$ HEA alloy using transport, magnetization, specific heat, and muon spin rotation/relaxation ($\mu$SR) measurements. Bulk superconductivity with transition temperature $T_{C}$ = 5.7 K confirmed by magnetization, resistivity, and heat capacity measurements. Zero-field $\mu$SR measurement shows that the superconducting state preserves time-reversal symmetry, and transverse-field measurements of the superfluid density are well described by an isotropic s-wave model.
\end{abstract}
\maketitle
\section{Introduction}
High entropy alloys (HEA) are a new class of multi-component material with outstanding mechanical, thermal, physical, and chemical properties \cite{metals, alloy}. High entropy alloys are typically a solid solution containing more than five principal elements \cite{2004Yeh,2004,cantor}. Configurations entropy (S) is also used as a parameter to define high entropy alloys. It should be comparable to $\Delta S_{mix}$ $>$ 1.5 $R$ for five-element alloy \cite{1.5}. The high entropy plays a vital role in the stabilization of the disordered solid solution phase \cite{prl}. It forms single-phase solid solutions with simple lattice despite having multiple components and complex chemistry. These disorder alloys have a broad spectrum of applications, such as anti-oxidation, anti-corrosion, thermoelectric, soft magnet, and radiation tolerance materials \cite{anti-oxy,anti-corr,radi-tol} over conventional alloys. In 2014, another fascinating electronic property, superconductivity, was observed first time in Ta$_{34}$Nb$_{33}$Hf$_{8}$Zr$_{14}$Ti$_{11}$ HEA with a transition temperature of 7.3 K \cite{prl}. It extends the possible applications of HEAs beyond their structural and mechanical properties-related applications. Since then, few other HEA or medium entropy alloy superconductors have been reported, which crystallized in BCC, FCC, CsCl, hexagonal, and tetragonal crystal structure \cite {alpha Mn,FCC,Hexagonal,Hexa,tetragonal}. To date, the maximum T$_{c}$ was achieved in (ScZrNbTa)$_{0.65}$(RhPd)$_{0.35}$ HEA with CsCl lattice-type crystal structure \cite{hightc}. Most of the effort on superconducting HEA is focused on discovering new superconducting HEAs families and enhancing superconducting transition temperature using the different combinations of 3$d$, 4$d$ and 5$d$ elements. Its multi-component nature offers the tunability of superconducting parameters and crystal structure. Some of these alloys show exotic superconducting properties such as retention of superconducting properties in very high-pressure \cite{pressure hea}, Debye temperature in elemental range \cite{alpha Mn,Hexa,muon hea} and phonon mediated superconducting similar to binary and ternary superconducting alloys despite its highly disordered nature, where the occurrence of regular phonon mode is unlikely. Understanding of superconductivity in high-entropy alloys remains a challenge, and the unavailability of microscopic measurements on superconducting HEA makes it difficult to understand the superconducting pairing mechanism of these disordered alloys \cite{muon hea}.
In this paper, we report the synthesis and detailed investigation of superconducting HEA Nb$_{0.6}$Re$_{0.1}$Zr$_{0.1}$Hf$_{0.1}$Ti$_{0.1}$ using bulk measurements of magnetization, heat capacity, and resistivity along with microscopic probe of the vortex lattice and superconducting ground state with muon spin spectroscopy. These measurements confirmed bulk type-II superconductivity in Nb$_{0.6}$Re$_{0.1}$Zr$_{0.1}$Hf$_{0.1}$Ti$_{0.1}$ having superconducting transition temperature T$_{C}$ $\approx$ 5.7 K. Superconducting ground state can be described as a nodeless isotropic gap with a moderate coupling strength and preserved time-reversal symmetry.
 
\section{Experimental Details}
The standard arc melting technique was used to synthesize a polycrystalline sample of Nb$_{0.60}$Re$_{0.10}$Zr$_{0.10}$Hf$_{0.10}$Ti$_{0.10}$. Elements Re (99.99$\%$), Nb (99.99$\%$), Hf (99.99$\%$), Zr (99.99$\%$), and Ti (99.99$\%$) with high purity were taken in the stoichiometric ratio and loaded in the cavity of the copper hearth. The chamber was filled with argon, and the cooling of the copper hearth was provided with continuous water flow. Firstly, we melted the titanium getter to absorb the oxygen, if any, present in the chamber and then melted the sample several times to ensure homogeneity. After several melting, the sample was shiny, and the weight loss was negligible. Sample composition was confirmed by energy dispersive X-ray spectrometer, and phase purity was confirmed by X-ray diffractometer (XRD) by using X'pert PANalytical diffractometer where Cu-K$_{\alpha}$ ($\lambda$ = 1.5405 \text{\AA}) radiation was used. Bulk superconductivity was confirmed by magnetization measurement using a superconducting quantum interference device (SQUID) from Quantum Design. Four probe and two tau methods were adopted to perform resistivity and specific heat, respectively using Quantum Design Physical Property Measurement System (PPMS). In order to investigate microscopic superconducting ground state and vortex lattice, the MuSR spectrometer was used at ISIS Neutron and Muon Source, STFC Rutherford Appleton Laboratory, United Kingdom, in zero field and transverse fields configurations.\\
\begin{figure} %{r}{0.5\textwidth}
\centering
\includegraphics[width=1.0\columnwidth, origin=b]{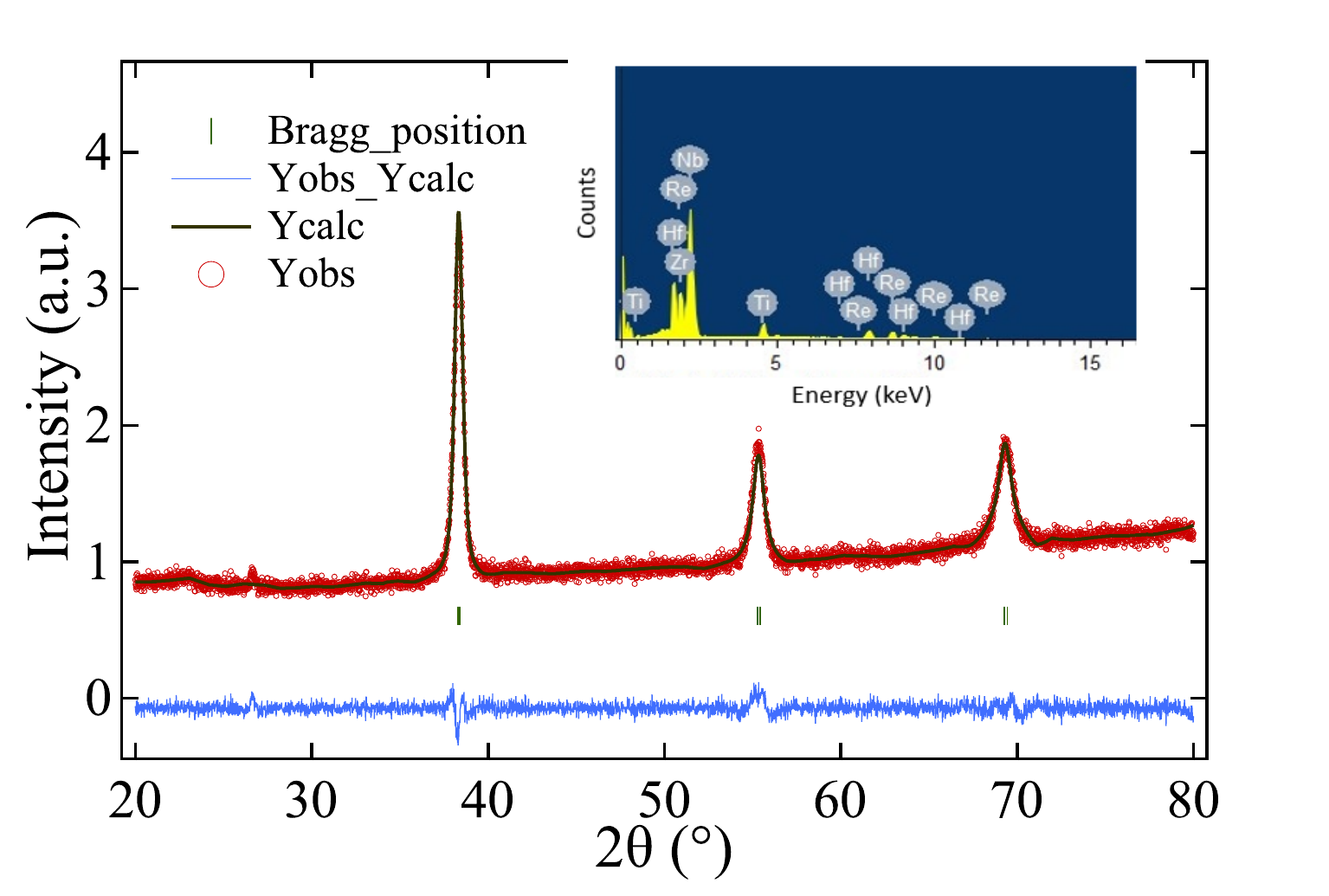}
\caption{ Powder X-ray diffraction pattern collected at room temperature by using CuK$_{\alpha}$ radiation and EDAX spectra shown in inset.}
\centering
\label{Fig1}
\end{figure}

\section{Results and Discussion}
\subsection{Structural characterization}
The X-ray data from the sample, at ambient conditions, was collected using the PANalytical X'Pert Pro instrument. The observed diffracted pattern was indexed with Full Prof software, and the refined pattern is displayed in \figref{Fig1}. It crystallized in centrosymmetric a BCC, space group I\textit{m-3m}, structure with unit cell parameter a = b = c = 3.320184(7) {\AA}. The broadness of the peaks may be due to the disorder created by different radii of elements within the sample \cite{prl}. The EDAX was performed at five different positions (one spectra was shown in the inset of \figref{Fig1}) on HEA sample, which confirms the homogeneity and the averaged composition is found out to be Nb$_{60}$Re$_{10}$Zr$_{10}$Hf$_{10}$Ti$_{10}$.
\begin{figure} %{r}{0.5\textwidth}
\centering
\includegraphics[width=1.0\columnwidth, origin=b]{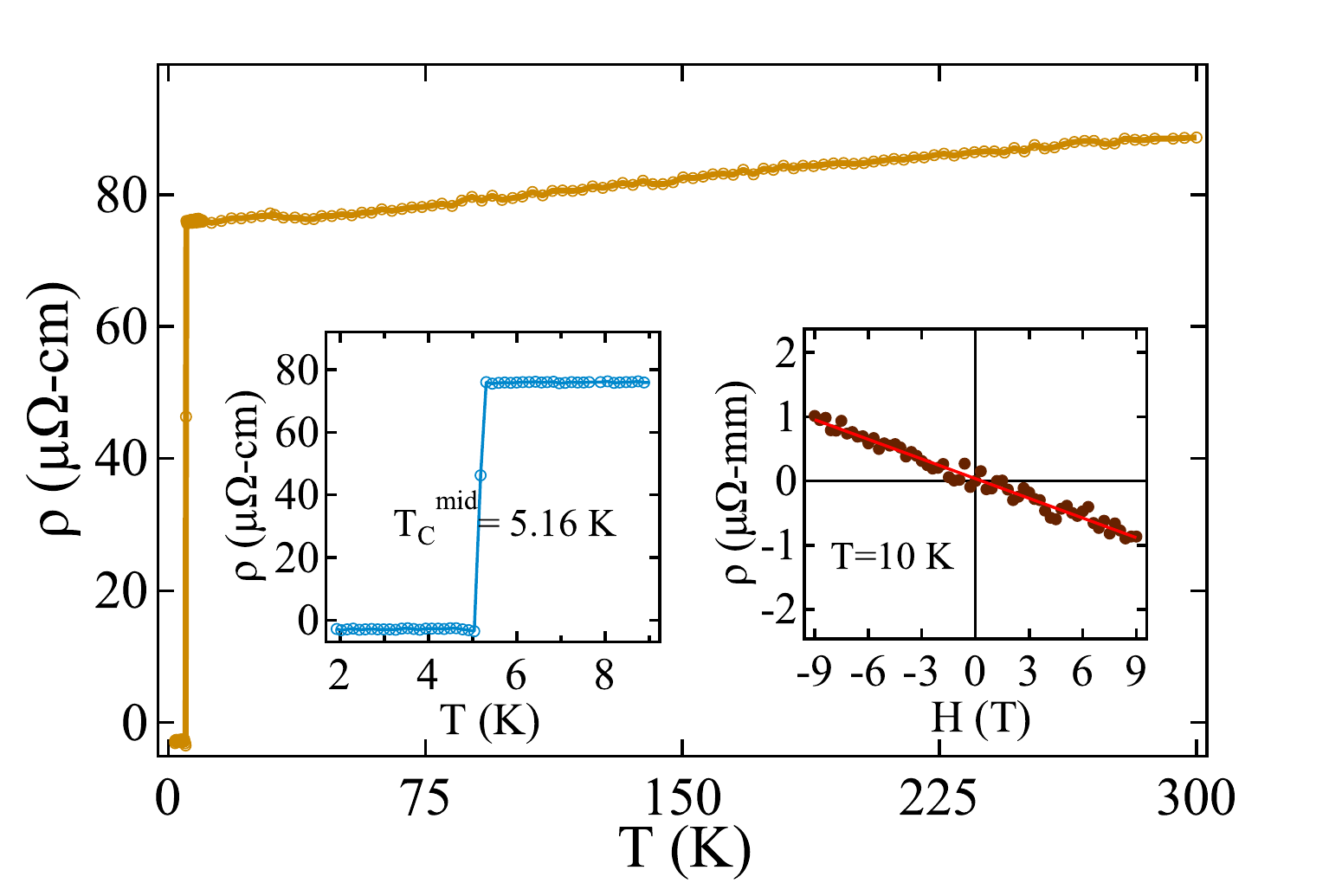}
\caption{ The temperature dependent resistivity in the range of 1.9 $K$-300 $K$ and the left inset shows the sharp transition at 5.2(1) $K$. the right inset shows field dependent resistivity at 10 $K$ in the range $\pm$ 9 $T$ and slope of resistivity vs magnetic field is negative.}
\centering
\label{Fig2}
\end{figure}
 \begin{figure*} %{r}{0.5\textwidth}
\includegraphics[width=2.0\columnwidth, origin=b]{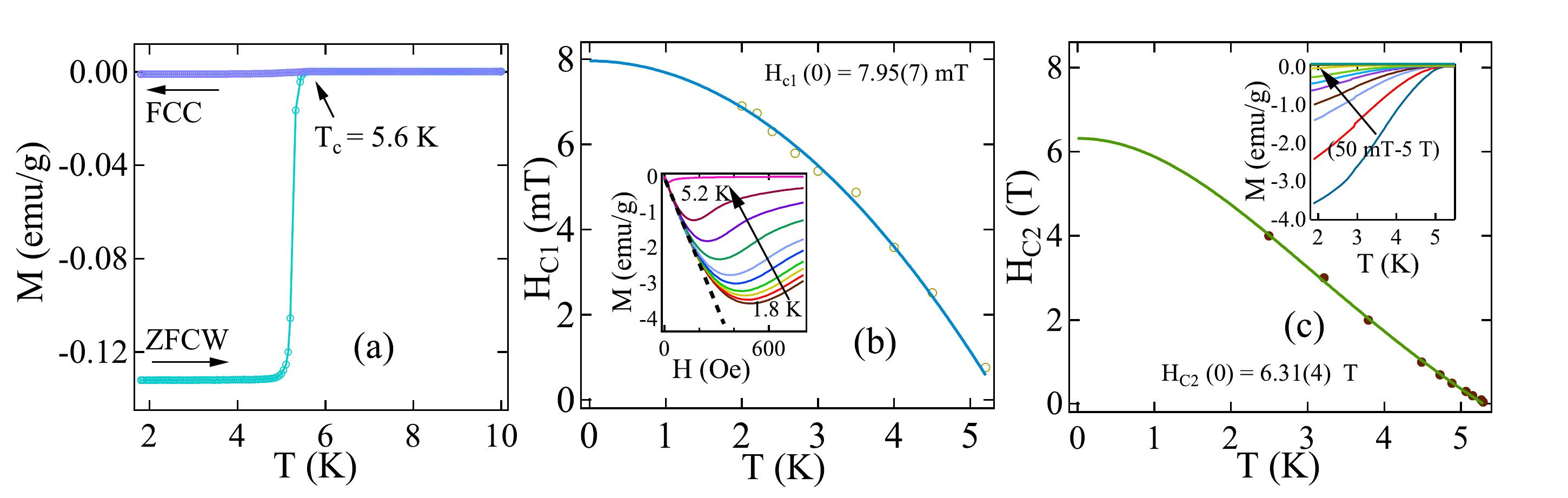}
\caption{(a) Temperature dependent of the DC magnetic moment in both ZFCW and FCC mode under the  applied field of 1 $mT$ (left).(b) Temperature dependent of lower critical field curved fitted with Gingburg-Landau equation (eqn. \ref{Hc1}) and inset of (b) shows the field dependent magnetization within temperature range from 1.8 $K$ to 5.2 $K$ (middle). (c) Temperature dependence data of the upper critical field fitted with the Gingburg-Landau equation (eqn.\ref{hc2}) and the inset of (c) magnetization versus temperature within range of field 50 $mT$ to 5 $T$ is shown(right).}
\label{Fig3}
\end{figure*}
\subsection {Electrical resistivity}
The electrical transport study of Nb$_{0.60}$Re$_{0.10}$Zr$_{0.10}$Hf$_{0.10}$Ti$_{0.10}$ and nature was revealed by temperature dependent resistivity measurement between temperature range 1.9 $K$ to 300 $K$ without applying any external field. The inset of Fig. \ref{Fig2} is displayed the sharp transition $T_{C}^{onset}$ at 5.30(5) $K$ and true zero resistive value $T_{C}^{base}$ = 5.03(5) $K$. We take the mid-value of the normal state to zero resistive states as the transition temperature $T_{C}^{mid}$ = 5.16(5) $K$. Above the transition temperature, the normal state resistivity was found to be almost temperature-independent and is shown in Fig \ref{Fig2}. The residual resistivity ratio (RRR) ($\rho_{300}$/$\rho_{10}$) is found out to be 1.2 which is comparable to other HEA superconductors \cite{muon hea,TNZHT,Cava}.
We have also measured Hall resistivity to study the nature of charge carrier and density in a normal state of the sample. The field-dependent hall resistivity measurement was done at 10 $K$, which is shown in the inset of Fig. \ref{Fig2}. The observed $\rho$ (H) data was linear fit within $\pm$ 9 $T$ magnetic field, and the slope provides the hall coefficient R$_{H}$ = - 1.08(2)$\times$10$^{-10}$ $\ohm m T^{-1}$. The negative sign of R$_{H}$ indicates that the electrons are the charge carriers. Using the expression R$_{H}$ = - 1/$ne$, where $n$ and $e$ are the carrier density and electronic charge, respectively. This expression yields the charge carrier density $n$ = 6.1(1) $\times$ 10$^{28}$ $m^{-3}$, which is comparable to the non-centrosymmetric (HfNb)$_{0.10}$(MoReRu)$_{0.90}$ and (ZrNb)$_{0.10}$(MoReRu)$_{0.90}$ high entropy alloy superconductors \cite{muon hea}.

\subsection{Magnetization}
We have performed temperature-dependent DC magnetization in zero-field cooled warming (ZFCW) and field-cooled cooling (FCC) mode under 1 $mT$ external magnetic field. A clear indication of diamagnetic signal was observed corresponding to the onset of superconducting transition temperature at $T_{C}$ = 5.60 (4) $K$ and is shown in Fig. \ref{Fig3}(a). The FCC data indicates the strong pinning nature in Nb$_{0.60}$Re$_{0.10}$Zr$_{0.10}$Hf$_{0.10}$Ti$_{0.10}$ alloy. Lower critical field $H_{C1}$ was determined using field-dependent magnetization data within a range of temperature from 1.8 K up to $T_{C}$ as shown inset of Fig. \ref{Fig3}(b). Up to a certain value of field, magnetization varies linearly and the point from which it starts deviating from the linearity or the Meissner line is taken as critical field for a particular temperature. The lower critical field value $H_{C1}$ at 0 $K$ was estimated after fitting the data points using  Ginzburg-Landau expression 
\begin{equation}
H_{C1}(T)=H_{C1}(0)\left(1-\left(\frac{T}{T_{C}}\right)^{2}\right) .
\label{Hc1}
\end{equation}
The extrapolation of $H_{C1} (T)$ to 0 $K$ yields $H_{C1}$(0) = 7.95(7) $mT$.\\
In order to estimate the upper critical field, temperature-dependent magnetization was carried out with varying magnetic fields. The onset of superconducting transition was considered as the criteria for the upper critical field, and transition temperature decreased with an increase of an applied external magnetic field, as shown in Fig. \ref{Fig3}(c). $H_{C2}$ versus temperature shows a linear behaviour near $T_{C}$ and can be described with the Ginzburg-Landau equation. 
\begin{equation}
H_{C2}(T) = H_{C2}(0)\left(\frac{(1-t^{2})}{(1+t^2)}\right). 
\label{hc2}
\end{equation}
\\
where $t = T/T_{C}$ is the reduced temperature, the fitting reveals the upper critical value at 0 $K$ is 6.31(4) $T$. The Gingburg-Landau coherence length parameter can be evaluated using the value of $H_{C2}$(0) according to the expression
\begin{equation}
H_{C2}(0) = \frac{\Phi_{0}}{2\pi\xi_{GL}^{2}}
\label{eqn3:up}
\end{equation}

where $\Phi_0$ is the flux quantum ($\Phi_0$ = 2.07$\times$10$^{-15}$T m$^2$) and substituting the $H_{C2}$(0) = 6.31(4) $T$, we get coherence length  $\xi_{GL}$(0) = 7.2(1) $nm$. 
The magnetic penetration depth $\lambda_{GL}$(0) depends on the lower critical field $H_{C1}$(0) and $\xi_{GL}$(0) according to the expression
\begin{equation}
H_{C1}(0) = \frac{\Phi_{0}}{4\pi\lambda_{GL}^2(0)}\left(\mathrm{ln}\frac{\lambda_{GL}(0)}{\xi_{GL}(0)}+0.12\right)   
\label{eqn6:ld}
\end{equation}
After substituting the calculated parameters $H_{C1}$(0) = 7.95(7) $mT$ and $\xi_{GL}$(0) = 7.2(1) $nm$, we obtained $\lambda_{GL}$(0) = 279(2) $nm$.
The Gingburg-Landau parameter $\kappa_{GL}$ differentiate the type of superconductivity and expressed as 
 $\kappa_{GL}$=$\frac{\lambda_{GL}(0)}{\xi_{GL}(0)}$.
 The $\xi_{GL}$ = 7.2(1) $nm$ and $\lambda_{GL}$(0) = 279(2) $nm$, we obtained $\kappa_{GL}$ = 39(1) >> ${\frac{1}{\sqrt{2}}}$ indicate that the Nb$_{60}$Re$_{10}$Zr$_{10}$Hf$_{10}$Ti$_{10}$ HEA is a strong type II superconductor.
In type II superconductors, the applied magnetic field is responsible for breaking the Cooper pair by two types of mechanism; orbital and Pauli limiting effects (briefly described in ref. \cite{Dsingh}).
 \begin{figure*} %{r}{0.5\textwidth}
\includegraphics[width=2.0\columnwidth, origin=b]{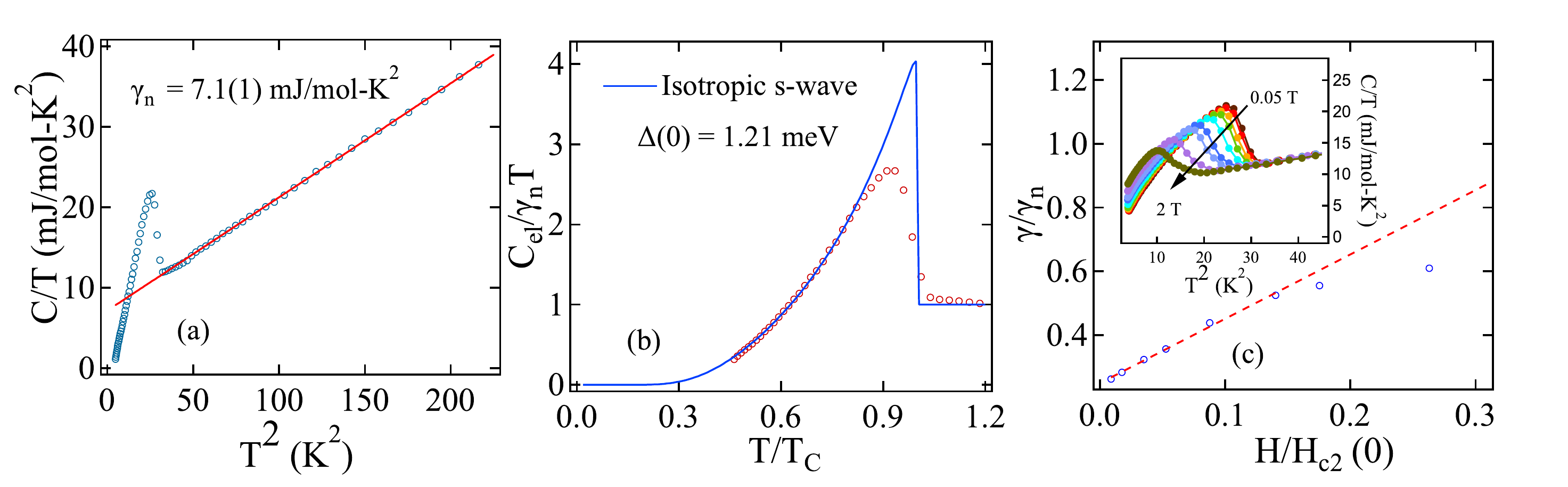}
\caption{(a) $C/T$ vs $T^{2}$ curve above $T_{C}$ fitted by equation \ref{eqn8:Cel}. (b) The observed data shown by red square circle was fitted with BCS isotropic gap (blue line) equation. (c) Normalised $\gamma$ of superconducting state is plotted with normalised field. Dashed red curve represents the linear behavior at lower values and indicate the $s$-wave superconductivity.}
\label{Fig4}
\end{figure*}
For BCS superconductor, the orbital limiting field $H_{C2}^{orb}$(0) is calculated via Werthamar-Helfand-Hohenberg (WHH) model
\begin{equation}
H_{C2}^{orbital}(0) = -\alpha   T_{C}\left.\frac{dH_{C2}(T)}{dT}\right|_{T=T_{C}}
\label{eqn4:whh}
\end{equation}
 where $\alpha$ can take 0.693 value in dirty limit. The ratio of BCS coherence length and electronic mean free path value  $\xi_{0}$/l >1 [detailed in section(f)] and suggest the dirty limit superconductivity in Nb$_{0.60}$Re$_{0.10}$Zr$_{0.10}$Hf$_{0.10}$Ti$_{0.10}$. The slope of temperature dependent upper critical field  ${\frac{dH_{C2}}{dT}}$ at $T=T_{C}$ is estimated to be  -1.08(3) $T$, which yields $H_{C2}^{orb}$ = 4.2(1) $T$, and another effect, Pauli limiting field $H_{P}$ can be addressed within the BCS theory using the expression H$_{c2}^{P}$ = 1.84 $T_{C}$ and the estimate value of $H_{C2}^{P}$ = 10.3 (1) $T$.\\ In order to measure the relative strength of orbital and Pauli limiting field, Maki parameter is calculated and expressed as $\alpha_{M} = \sqrt{2}H_{C2}^{orb}(0)/H_{C2}^{p}(0)$. The value obtained for $\alpha_{M}$= 0.57, suggesting the small influence of paramagnetic effect and mainly orbital limiting field is responsible for pair breaking.\\

\subsection{Specific heat}
Specific heat measurement was performed on Nb$_{0.60}$Re$_{0.10}$Zr$_{0.10}$Hf$_{0.10}$Ti$_{0.10}$ alloy with applying external field and in zero fields. The sample was mounted on the platform with Apiezon N grease to establish good thermal contact. A pronounced jump in heat capacity was observed at transition temperature $T_{C}$ at 5.4(1) $K$ as shown in Fig. \ref{Fig4}(a). The $T_{C}$ value estimated by specific heat measurement is comparable with resistivity and magnetization. The specific heat data in the normal state (above $T_{C}$) was well fitted with the following equation:
\begin{equation}
\frac{C}{T} = \gamma_{n} + \beta_{3}T^{2}
\end{equation}
This fitting provides the electronic parameter (Sommerfeld coefficient) $\gamma_n$ = 7.1(1) $mJmol^{-1}$K$^{-2}$ and the lattice contribution parameter to the specific heat, $\beta_3$ = 0.140(2) $mJmol^{-1}$K$^{-4}$. Both the parameters, $\gamma_n$ and $\beta_3$ have paramount importance in calculating the density of states ($D_{c} (E_{F})$), Debye temperature ($\theta_{D}$), and electron-phonon coupling constant ($\lambda_{e-p}$). The following quantities are calculated from the expressions provided in \cite{Dsingh} and the estimated values are: $D_{c} (E_{F})$ = 3.01(4) $\frac{states}{eV f.u}$, $\theta_{D}$ = 240(1) $K$, and $\lambda_{e-p}$ = 0.70. The value of electron phonon coupling strength $\lambda_{e-p}$ is comparable with other High entropy alloy superconductors \cite{tetragonal,poly}, which suggests moderately coupled superconductivity in Nb$_{0.60}$Re$_{0.10}$Zr$_{0.10}$Hf$_{0.10}$Ti$_{0.10}$.\\
To further reveal the superconducting ground state behavior, electronic specific heat $C_{el}$ in the superconducting state can be estimated by excluding the phonon contribution from the total specific heat $C(T)$ by the expression $C_{el}$ = $C(T)$ - $\beta_{3}T^3$.
For a single gap BCS superconductor, the normalized entropy in the superconducting state is given by
\begin{equation}
\frac{S}{\gamma_{n}T_{C}} = -\frac{6}{\pi^2}\left(\frac{\Delta(0)}{k_{B}T_{C}}\right)\int_{0}^{\infty}[ \textit{f}\ln(f)+(1-f)\ln(1-f)]dy \\
\label{eqn7:s}
\end{equation}
\\
where $\textit{f}$($\xi$) = [exp($\textit{E}$($\xi$)/$k_{B}T$)+1]$^{-1}$ is the Fermi function, $\textit{E}$($\xi$) = $\sqrt{\xi^{2}+\Delta^{2}(t)}$, where E($ \xi $) is the energy of the normal electrons relative to the Fermi energy, $\textit{y}$ = $\xi/\Delta(0)$, $\mathit{t = T/T_{C}}$ and $\Delta(t)$ = tanh[1.82(1.018(($\mathit{1/t}$)-1))$^{0.51}$] is the BCS approximation for the energy gap. The normalized electronic specific heat below $T_{C}$ is related to the first derivative of normalized entropy as \\
\begin{equation}
\frac{C_{el}}{\gamma_{n}T_{C}} = t\frac{d(S/\gamma_{n}T_{C})}{dt} \\
\label{eqn8:Cel}
\end{equation}
\\
The fitting provides the superconducting gap value $\frac{\Delta(0)}{k_B T_C}$ = 2.57, which is very high in magnitude with respect to the BCS predicted value and slightly more than the equimolar Nb$_{20}$Re$_{20}$Zr$_{20}$Hf$_{20}$Ti$_{20}$ HEA \cite{maneesha}. The electronic heat capacity data together with electron phonon coupling strength provide the evidence of isotropic $s$ - wave moderately coupled superconductivity in Nb$_{60}$Re$_{10}$Zr$_{10}$Hf$_{10}$Ti$_{10}$ . 
To further unveil the superconducting gap structure, we carried out field-dependent specific heat measurements. We first calculated the electronic specific heat with respective fields by subtracting the corresponding phononic part ($\beta_3$T$^{3}$) from total specific heat (C). The field-dependent electronic specific heat in the superconducting state was fitted with the expression \cite{low BCS}
\begin{equation}
\frac{C_{el}}{T} = \gamma + \frac{A}{T}\exp\left(\frac{-bT_{c}}{T}\right)
\end{equation}
here A is material dependent quantity, and b = $\frac{\Delta(0)}{k_{B}T_{c}}$ superconducting gap. Fig. \ref{Fig4}(c) shows the normalized $\gamma$ rises linearly with applied external magnetic field H and both the parameters $\gamma$ and H are normalized by $\gamma_{n}$ and the upper critical field $H_{C2}$(0), respectively. The linear increase in $\gamma$ is because the quasiparticle excitation is confined within the vortex cores and a number of vortices vary with field $H$ for an isotropic superconducting gap structure \cite{low BCS,gama H}. Thus, the linear behavior of $\gamma (H)$ proportional to field $H$ at low fields indicates the fully isotropic $s$-wave superconducting gap. In the presence of nodes, the $\gamma (H)$ is expected to vary as square root to the applied field.\\
\subsection{$\mu$SR Spectroscopy}
Muon spin relaxation in zero fields and muon spin rotation in transverse field measurements were done on a MuSR spectrometer ISIS facility at Rutherford Appleton Laboratory, U. K. Forty pulses of spin-polarized muons enter MuSR in each second, which are implanted into the sample and these muons sits at the interstitial positions in the powder sample. The muon spin precess with Larmor frequency, decays into a positron and two neutrinos. The positron decay into the initial muon spin direction, is recorded through 64 scintillator detectors, and these detectors are positioned in a circular design around the sample chamber. In zero field muon spin relaxation measurement, two sets of the detectors were positioned around the sample, while in transverse field muon spin rotation, the detectors were positioned in four configurations, and each configuration consist of 16 detectors which are named as forward, backward, top and bottom. The time domain asymmetry spectra were calculated using the expression
A(t) = $\left(\frac{F(t)- \alpha B(t)}{F(t)+ \alpha B(t)}\right)$
and $\alpha$ is the relative efficiency of counts between any two of the four detectors \cite{aidy}.
This asymmetry function gives information regarding muon spin polarization at a time scale. The powder sample was mixed with GE varnish and mounted on a silver holder. The benefits of using a silver holder are getting time-independent signal in ZF-$\mu$SR and a non-decaying oscillation in TF-$\mu$SR configurations. The low temperature was achieved using $^{3}$He sorption cryostat. In the ZF configuration, the sample was first cooled to the lowest temperature in the zero field, and then data was taken while warming without any applied field. The effect of the magnetic field at sample position due to the earth's magnetic field or from the neighboring instruments are cancelled by a set of correction coil. In TF configuration, an external magnetic field was applied perpendicular to the direction of muon spin. The sample was field cooled to create a flux line lattice and data were taken from base temperature to above the transition temperature. \\
\begin{figure} %{r}{0.5\textwidth}
\centering
\includegraphics[width=1.0\columnwidth, origin=b]{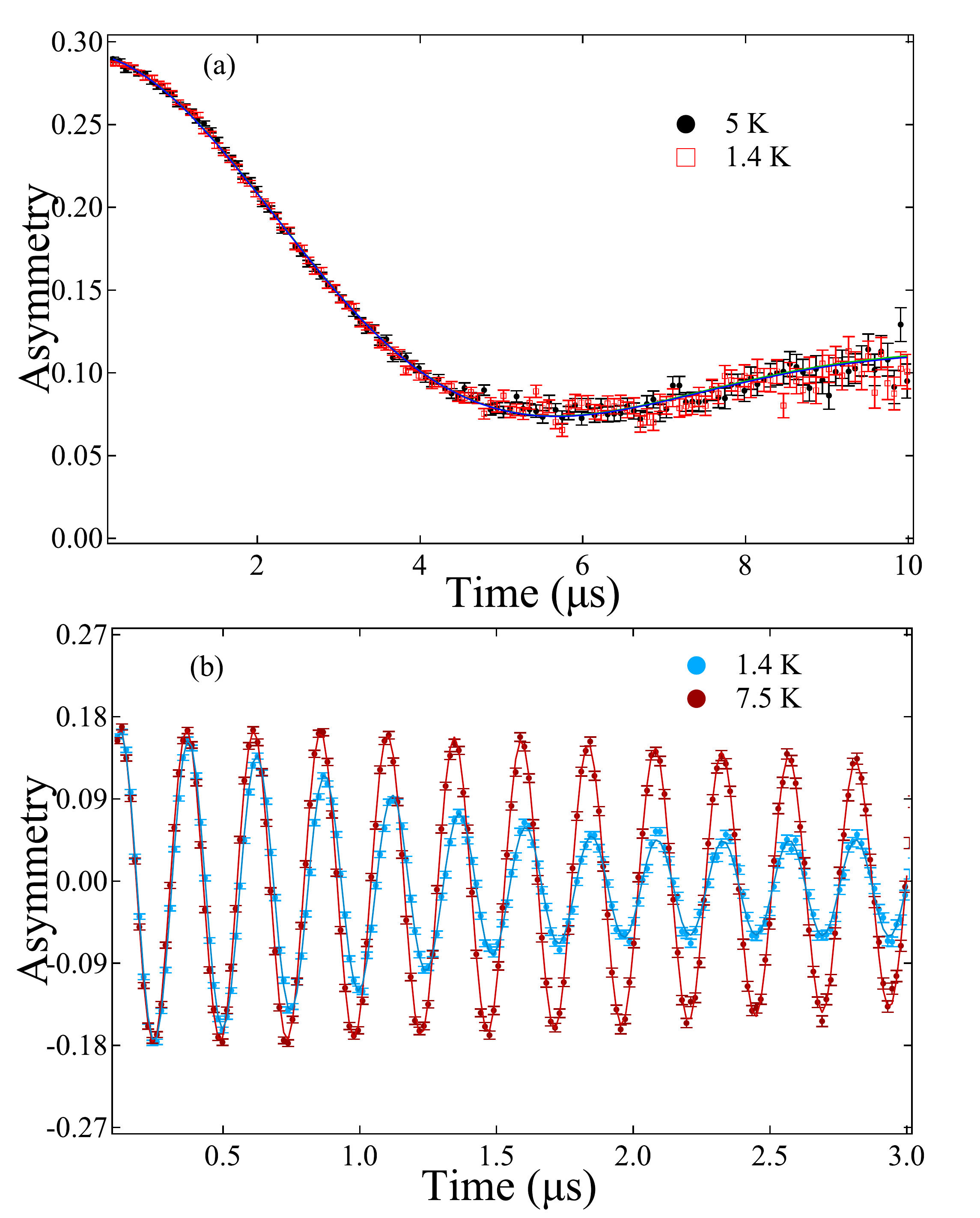}
\caption{ (a) Zero field $\mu$SR asymmetry spectra is shown from above and below T$_{C}$. The black line represents the fit to the Kubo-Toyabe equation. (b) Transverse field $\mu$SR asymmetry spectra collected at 1.39 $K$ and 7.5 $K$ under applied field of 30 $mT$. These red and blue lines are the fit to the data using equation \ref{eqn2:TF1}.}
\label{Fig5}
\end{figure}
\subsubsection{ZF-$\mu$SR Relaxation}
$\mu$SR measurements in zero field can be used to detect spontaneous magnetization associated with broken time-reversal symmetry in the superconducting state.
\begin{figure*} %{r}{0.5\textwidth}
\includegraphics[width=2.0\columnwidth, origin=b]{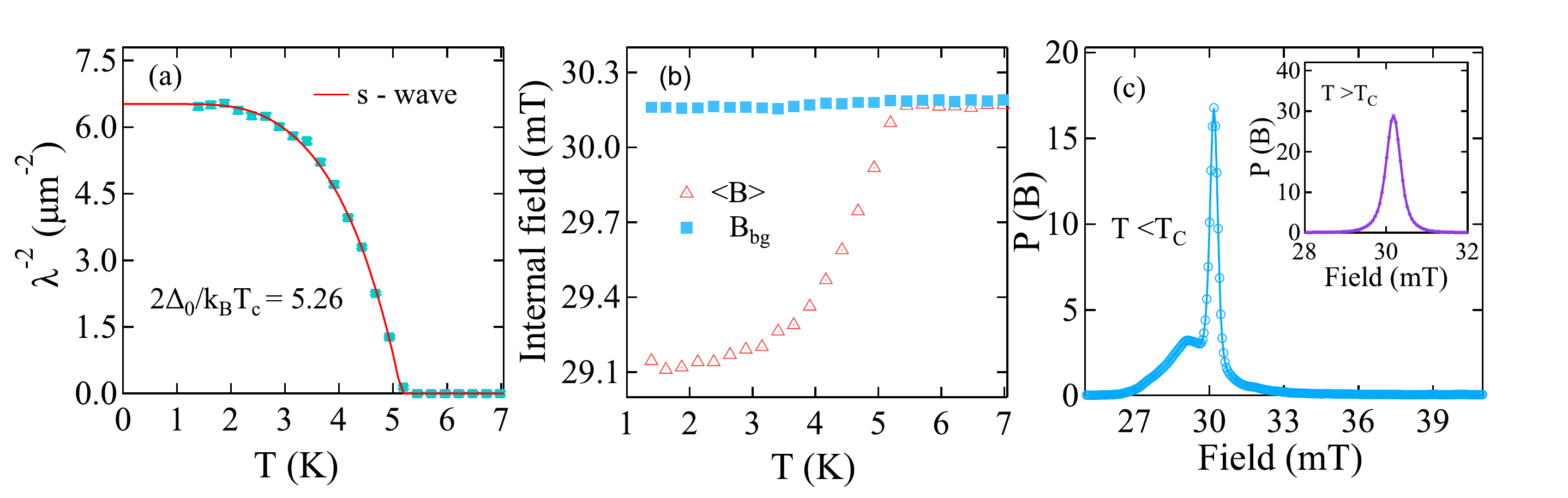}
\caption{(a) Temperature dependent variation of $\lambda^{-2}$ fitted well with isotropic $s$-wave BCS superconducting model. (b) Internal magnetic field distribution inside the sample with respect to temperature. Red data points are the observed field from the sample and blue data points are the background field contribution from the sample holder. (c) Field distribution in vortex state and normal state (inset of Fig. \ref{Fig6}(c).}
\label{Fig6}
\end{figure*}
The ZF-$\mu$SR is the best tool to investigate the minimal magnetic field without applying an external magnetic field. We collected the relaxation spectra above and far below the superconducting transition temperature, which is shown in Fig. \ref{Fig5}(a). The decaying relaxation behavior indicates that no oscillation present above and below $T_{C}$, and the relaxation comes from the randomly oriented nuclear moments. ZF-$\mu$SR relaxation data above and below $T_{C}$ were fitted very well with the Kubo-Toyabe equation, which is given as \cite{KT} \\
\begin{equation}
G_{\mathrm{KT}}(t) = \frac{1}{3}+\frac{2}{3}(1-\sigma^{2}_{\mathrm{ZF}}t^{2})\mathrm{exp}\left(\frac{-\sigma^{2}_{\mathrm{ZF}}t^{2}}{2}\right) ,
\label{eqn9:zf}
\end{equation} \\
where $\sigma_{ZF}$ is the muon spin relaxation rate. The asymmetry spectra can be explained with expression as  
\begin{equation}
A(t) = A_{1}G_{\mathrm{KT}}(t)\mathrm{exp}(-\Lambda t)+A_{\mathrm{BG}} 
\label{eqn10:tay}
\end{equation}
here A$_{1}$ is the sample asymmetry, A$_{BG}$ is the background asymmetry signal coming due to the stopping of muons in the sample holder, and $\Lambda$ is electronic relaxation rate. The fitting of \equref{eqn10:tay} above and below the T$_{C}$, yield no noticeable change of relaxation rate in the detection limit (10 $\mu$T) of instruments. The change corresponding to both the relaxation are found as $\sqrt \Delta /\gamma_{\mu}$ = 0.01 Oe and $\Lambda / \gamma_{\mu}$ = 0.04 Oe, which are outside the scope of detection for $\mu$SR instrument. Hence, the time reversal symmetry is preserve in Nb$_{60}$Re$_{10}$Zr$_{10}$Hf$_{10}$Ti$_{10}$.

\subsubsection{TF-$\mu$SR}
To detect the superconducting gap symmetry and the field distribution in the mixed state, TF-$\mu$SR measurements have been performed on Nb$_{0.60}$Re$_{0.10}$Zr$_{0.10}$Hf$_{0.10}$Ti$_{0.10}$ alloy. In this experiment, we first cool the sample in an applied magnetic field of 30mT, which is greater than the lower critical field but less than the $H_{c2}$ ($H_{c1}$ < $H^{ext}$ << $H_{C2}$) to create a well-ordered flux line lattice. Asymmetry spectra in TF-$\mu$SR were recorded in temperature range from 1.39 $K$ to 7.5 $K$ and only two spectra at 1.39 $K$ and above $T_{C}$ are displayed in Fig. \ref{Fig5}(b). The distribution of field in the entire temperature scale was extracted from the TF-$\mu$SR time spectra by the Maximum Entropy method (ME) \cite{ME}. Below the T$_{C}$, Fig. \ref{Fig6}(c) is showing the evidence of depolarization, and inset of Fig. \ref{Fig6}(c)) is clearly showing the homogeneous field above $T_{C}$. The variation of muon precession in asymmetry spectra was observed because of the interaction of muon spin with internal field distribution in a sample. Above $T_{C}$, muon spin relax due to the presence of randomly oriented nuclear dipolar moments, and below $T_{C}$, the relaxation arises from the inhomogeneous magnetic field due to the formation of flux line lattice in the mixed state plus the nuclear contribution. The asymmetry spectra of TF-$\mu$SR is well fitted in all temperature range by the Gaussian oscillatory decaying function, and it is described below as \cite{KT1,KT2} 

\begin{equation}
\begin{split}
A (t) = \sum_{i=1}^N A_{i}\exp\left(-\frac{1}{2}\sigma_i^2t^2\right)\cos(\gamma_\mu B_it+\phi)\\ + A_{bg}\cos(\gamma_\mu B_{bg}t+\phi),
\label{eqn2:TF1}
\end{split}
\end{equation}
where $A_{i}$ and $A_{bg}$ are the sample and sample holder related asymmetry signal respectively, $\phi$ is the initial phase offset and $\sigma$ is rate of Gaussian depolarization of muon spin. $B_{i}$ is mean field contribution of $i^{th}$ component of the Gaussian distribution, and $B_{bg}$ is the background field contribution coming from the sample holder.\\
Above $T_{C}$, there is a constant relaxation rate that occurs due to the random orientation of magnetic moment of nuclear spins and is temperature independent. The internal field dependence on temperature is shown in \figref{Fig6}(b), which shows a background magnetic field contribution that is constant for the temperature range measured. The internal magnetic field distribution in the mixed state denoted by $<$B$>$ is less than the applied fields for $T$ < $T_{C}$, which clearly shows the Meissner field expulsion in the superconducting state. At temperature ($T$ > $T_{C}$), it again recovers the applied field value and overlaps with the background signal. The total depolarization rate is not only from the flux lattice ($\sigma_{FLL}$) but also arises due to the nuclear magnetic moment ($\sigma_{n}$), which is temperature independent. $\sigma_{n}$ contribution was calculated from the above $T_{C}$ data points and $\sigma_{FLL}$ is estimated from subtracting the contribution of $\sigma_{n}$ from the total depolarization rate as $\sigma_{FLL}^{2} = \sigma^2-\sigma_{n}^2$, where $\sigma$ is the total depolarization rate. $\sigma_{FLL}$ is related to the London penetration depth via ($\sigma_{FLL}$ $\propto$ 1/$\lambda^{-2}$). For small applied fields where H << $H_{C2}$, the inverse square of the London penetration depth can be calculated via the formula: 
\begin{equation}
\frac{\sigma_{\mathrm{FLL}}^2(T)}{\gamma_{\mu}^2} = \frac{0.00371\Phi_{0}^2}{\lambda^{4}(T)}
\label{eqn13:sigmaH}
\end{equation}
where $\Phi_{0}$ is the magnetic flux quantum and $\gamma_{\mu}$/2$\pi$ = 135.5 $MHz/T$ is muon gyromagnetic ratio. Within London approximation, the calculated London penetration depth at 0 $K$ is $\lambda^{\mu}$ = 391(18) $nm$, which is bit higher than the calculated value of $\lambda$ = 279(2) $nm$ from magnetization.
The temperature dependence of penetration depth for $s$-wave BCS superconducting gap is given by the expression \cite{BCS L1,BCS L2,BCS L3}
\begin{equation}
\begin{split}
\frac{\lambda^{-2}(T)}{\lambda^{-2}(0)} = 1+2 \int_{\Delta(T)}^\infty(\frac{\partial f}{\partial E})\frac{E dE}{\sqrt{E^2-\Delta(T)^2}},
\label{eqn5:S-wave}
\end{split}
\end{equation}
Where $\lambda$(0) is the London penetration depth at 0 $K$, and $f$ = [1+$\exp$(E/k$_{B}$T)]$^{-1}$ is the Fermi function and $\Delta$ (T) is the BCS superconducting gap function which is defined by $\Delta$ (T) = $\Delta_{0}$tanh[1.82(1.018($\mathit{(T_{C}/T})$-1))$^{0.51}$]. The experimental data of $\lambda^{-2}$ (T) was best fitted with the BCS s-wave modal given by \equref{eqn5:S-wave} and the obtained T$_C$ = 5.2(2) $K$ is also in good agreement with the values measured from magnetization and specific heat measurements for Nb$_{60}$Re$_{10}$Zr$_{10}$Hf$_{10}$Ti$_{10}$. The ratio of superconducting gap at 0 $K$, $\Delta_0$/k$_B$T$_c$ = 2.65(5), which shows great deviation from the BCS value (1.73) and comparable to estimated value by specific heat measurement (2.57(8).\\
 \begin{figure}%{r}{0.5\textwidth}
\includegraphics[width=1.0\columnwidth, origin=b]{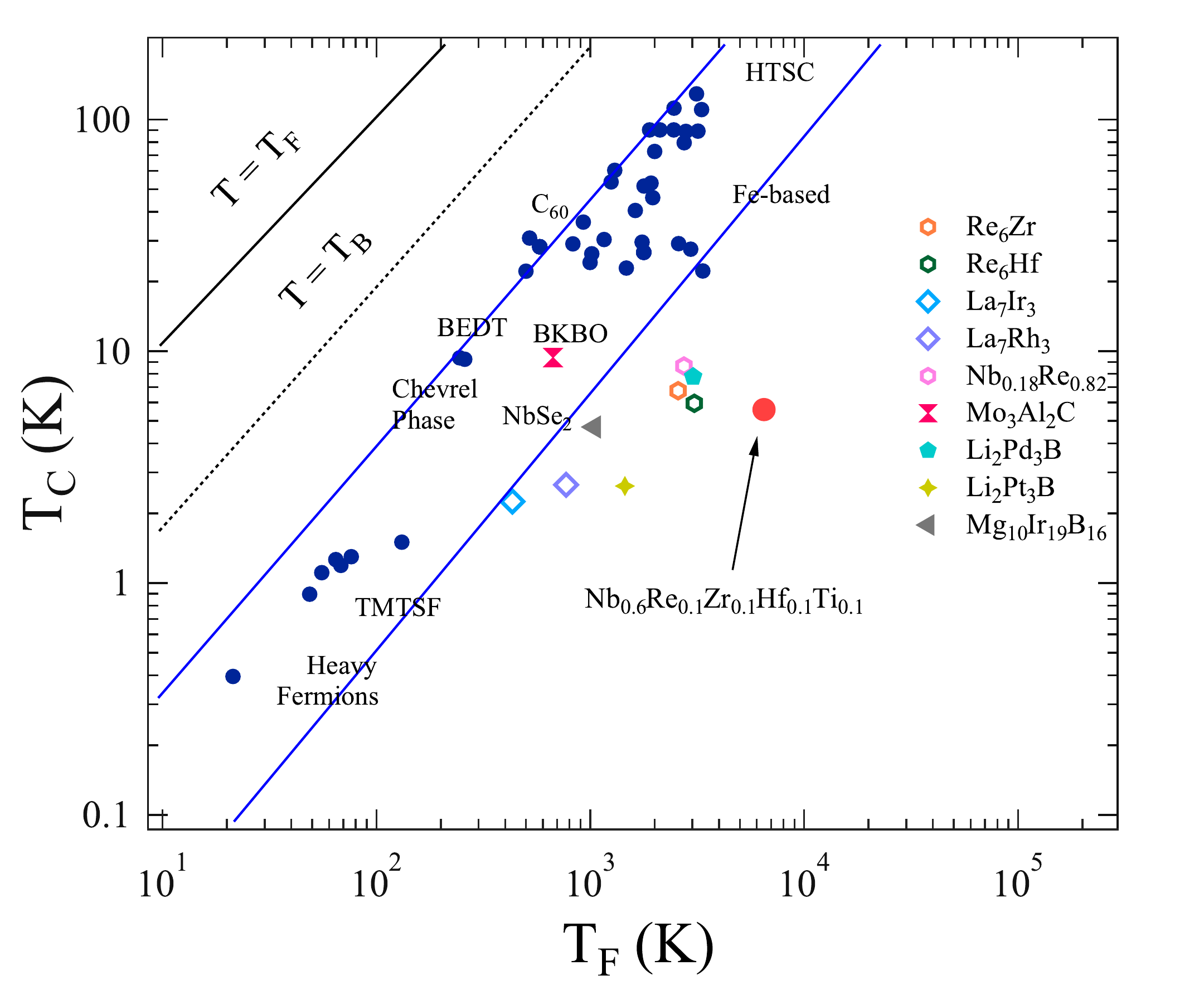}
\caption{Uemura plot representing the superconducting transition temperature versus Fermi temperature for different superconducting families. Two solid blue lines represents the unconventional band of superconductors and the solid red circle for Nb$_{60}$Re$_{10}$Zr$_{10}$Hf$_{10}$Ti$_{10}$ alloy away from the unconventional band.}
\label{Fig7}
\end{figure}
\subsection{Electronic property}
To supplement the observed result of experiment measurements, we have determined the electronic parameter such as the BCS coherence length, electronic mean free path and the Fermi velocity. We have used estimated parameter charge carrier density (measured from hall measurement) shown inset of Fig. \ref{Fig2}, Sommerfeld coefficient by normal heat capacity, and resistivity at transition temperature. The Sommerfeld coefficient is related with effective mass and carrier density of quasiparticle by the equation \cite{Thinkam}
\begin{equation}
\gamma_{n} = \left(\frac{\pi}{3}\right)^{2/3}\frac{k_{B}^{2}m^{*}n^{1/3}}{\hbar^{2}}
\label{eqn16:gf}
\end{equation}
where $k_{B}$, $n$ and $m^{*}$ are Boltzmann constant, carrier density and effective mass of quasiparticle, respectively. Using the values of Sommerfeld coefficient $\gamma_{n}$ =  7.1(1) $mJ-mol^{-1}$K$^{-2}$ and carrier density $n$ = 6.1(1)$\times {10}^{28} m^{-3}$, we estimated the effective mass of quasi-particle $m^{*}$ = 9.9(2) $m_{e}$.\\ 
The Fermi velocity is directly related to the carrier density and effective mass, which can be expressed as 
\begin{equation}
n = \frac{1}{3\pi^{2}}\left(\frac{m^{*}v_{\mathrm{f}}}{\hbar}\right)^{3} .
\label{eqn18:n}
\end{equation}
Using the estimated value of effective mass $m^{*}$ = 9.9 $m_{e}$ and carrier density $n$ = 6.1(1)$\times {10}^{28} m^{-3}$ of quasiparticle, we get the Fermi velocity $v_{F}$ = 1.5$\times$ 10$^{5}$ $ms^{-1}$.\\ 
The mean free path is dependent on Fermi velocity $v_{f}$, effective mass $m^{*}$ and the residual resistivity by the expression 
\begin{equation}
\textit{l} = \frac{3\pi^{2}{\hbar}^{3}}{e^{2}\rho_{0}m^{*2}v_{\mathrm{F}}^{2}}
\label{eqn17:le}
\end{equation}
The previous estimated value of Fermi velocity and effective mass with the residual resistivity $\rho_{0}$ = 76(1) $\mu \ohm-cm$, which yield mean free path $l$ = 9.6 \textup{\AA}.\\  
BCS coherence length can be estimated using Fermi velocity and transition temperature by the expression as
\begin{equation}
\xi_{0} = \frac{0.18{\hbar}{v_{F}}}{k_{B}T_{C}}
\label{eqn19:f}
\end{equation}
 Using the Fermi velocity parameter value and transition temperature $T_{C}$ = 5.6 $K$, we get the ratio of BCS coherence length to electronic mean free path $\xi(0)$/$l$ = 38 > 1, which strongly suggested that the sample is in the dirty limit of superconductivity. Other estimated parameters are listed in Table I.\\
The Fermi temperature can be expressed as;
 $k_{B}T_{F}$ = ${\frac{{\hbar}^{2}}{2m^{*}} (3{\pi}^{2}n)^{2/3}}$
where $k_{B}$, $m^{*}$ = 9.9(2) $m_{e}$, and $n$ = 6.1(1)$\times {10}^{28} m^{-3}$ are the Boltzmann constant, effective mass, and carrier density of quasi-particle, respectively. The expression gives Fermi temperature, T$_{F}$ ~ 6480(236) $K$.
Uemura $et$ $al$. \cite{U1,U2,U3} give a classification to distinguish the conventional and unconventional nature of superconductor by the ratio of T$_{C}$/T$_F$. The range of unconventional superconductor is 0.01 $\leq$ T$_{C}$/T$_F$ $\leq$ 0.1, and rest of the range represents the conventional superconductor. The ratio of T$_{C}$/T$_F$ = 0.00086 as shown in Fig. \ref{Fig7}, which places Nb$_{60}$Re$_{10}$Zr$_{10}$Hf$_{10}$Ti$_{10}$ away from the  unconventional superconductor line, and categories the Nb$_{60}$Re$_{10}$Zr$_{10}$Hf$_{10}$Ti$_{10}$ as a conventional superconductor.
\begin{table}[h!]
\caption{Superconducting and normal state parameters of Nb$_{60}$Re$_{10}$Zr$_{10}$Hf$_{10}$Ti$_{10}$}
\begin{center}
\begin{tabular}[b]{lccc}\hline
PARAMETERS& UNIT& Nb$_{0.6}$Re$_{0.1}$Zr$_{0.1}$Hf$_{0.1}$Ti$_{0.1}$\\
\hline
\\[0.5ex]                                  
$T_{C}$& K& 5.60(4)\\             
$H_{C1}(0)$& mT& 7.95(7)\\                  H$_{C2}^{mag}$(0)& T& 6.31(1)\\
$H_{C2}^{P}(0)$& T& 10.3(1)\\
$\xi_{GL}$& \text{$nm$}& 7.2(1)\\
$\lambda_{GL}$& \text{$nm$}& 279(2)\\
$\lambda(0)^\mu$& \text{$nm$}& 391(18)\\
$k_{GL}$& &39(1)\\
$\Delta(0)/k_{B}T_{C}$(specific heat)&  &2.57(8)\\
$\Delta(0)/k_{B}T_{C}$(muon)&  &2.65(5)\\
$m^{*}/m_{e}$& & 9.9(2)\\ 
$\xi_{0}/l_{e}$&   & 38(5)\\
$v_{F}$& ${10}^{5}m s^{-1}$& 1.50(4)\\
$n_s$& 10$^{28}$m$^{-3}$& 6.1(1)\\
$T_{F}$& K& 6480(236)
\\[0.5ex]
\hline\hline
\end{tabular}
\par\medskip\footnotesize
\end{center}
\end{table}
 
\section{Conclusion}
A detailed investigation of the superconducting and normal state properties of the superconducting HEA Nb$_{60}$Re$_{10}$Zr$_{10}$Hf$_{10}$Ti$_{10}$ has been conducted, using detailed magnetization, electrical resistivity, heat capacity, and $\mu$SR measurements. It is a type-II superconductor with a nodeless, isotropic superconducting gap and is well described by the BCS theory in the dirty limit. In the superconducting ground state, time-reversal symmetry has been preserved in the detection limit of $\mu$SR experiments. Combined results from all of the experimental techniques used to calculate the electronic properties and Umera plot. It is shown that superconductivity in Nb$_{60}$Re$_{10}$Zr$_{10}$Hf$_{10}$Ti$_{10}$ is in the dirty limit, and the Uemura plot placed it in the region of conventional superconductor. To understand the superconducting properties of HEAs, it is essential to search and perform a microscopic study of new superconducting HEAs.

\section{Acknowledgments}
Kapil Motla acknowledges the CSIR  funding agency (Award no; 09/1020(0123)/2017-EMR-I), Council of Scientific and Industrial Research (CSIR) Government of India for providing SRF fellowship. R.~P.~S.\ acknowledge Science and Engineering Research Board, Government of India for the Core Research Grant CRG/2019/001028. We thank ISIS, STFC, UK for the beamtime to perform the $\mu$SR experiments.

\end{document}